\documentclass[lettersize,journal]{IEEEtran}
\usepackage{amsmath,amsfonts}
\usepackage{algorithmic}
\usepackage{algorithm}
\usepackage{array}
\usepackage[caption=false,font=normalsize,labelfont=sf,textfont=sf]{subfig}
\usepackage{textcomp}
\usepackage{stfloats}
\usepackage{url}
\usepackage{verbatim}
\usepackage{graphicx}
\usepackage[numbers,sort]{natbib}
\usepackage{enumerate}
\usepackage{xspace}
\usepackage[utf8]{inputenc}
\usepackage{fontenc}
\usepackage{bm}
\usepackage{bbm}
\usepackage{gensymb}
\usepackage{xfrac}
\usepackage{textcomp}  
\usepackage[]{acronym}
\usepackage{colortbl}
\usepackage{xcolor}
\usepackage{relsize}
\usepackage{stfloats} 

\title{DYRECT Computed Tomography: DYnamic Reconstruction of Events on a Continuous Timescale}

\author{Wannes Goethals, Tom Bultreys, Steffen Berg, Matthieu~N.~Boone, Jan Aelterman
\thanks{W. Goethals (W.G.), T. Bultreys (T.B.), M.N. Boone (M.N.B.), and J. Aelterman (J.A.) are with UGCT at Ghent University.}
\thanks{W.G., M.N.B., and J.A. are affiliated to the Department of Physics and Astronomy, Radiation Physics, Ghent University,  Proeftuinstraat 86/N12, B-9000~Gent, Belgium. }
\thanks{W.G. and T.B. are with the Department of Geology, Pore-Scale Processes in Geomaterials, Ghent University,  Krijgslaan 281/S8, B-9000~Gent, Belgium}
\thanks{S. Berg is with Shell Global Solutions International B.V., Grasweg 31, 1031 HW Amsterdam, The Netherlands}
\thanks{Manuscript received October 23, 2024; revised January 16, 2024; accepted April 18, 2025.}}

\begin{document}

\maketitle

\begin{abstract}
Time-resolved high-resolution X-ray Computed Tomography (4D~$\mu$CT) is an imaging technique that offers insight into the evolution of dynamic processes inside materials that are opaque to visible light.
Conventional tomographic reconstruction techniques are based on constructing a sequence of 3D images from radiographic projections, recorded during time-frames that represent global sample states.
This frame-based approach limits the temporal resolution compared to dynamic radiography experiments, and it leads to an inflation of the amount of data. This results in costly post-processing computations to quantify the dynamic behaviour from the sequence of time-frames, hereby often ignoring the temporal correlations of the sample structure.
Our proposed 4D $\mu$CT reconstruction technique, named DYRECT, estimates individual attenuation evolution profiles for each position in the sample with time resolution down to the single projection level.
This leads to a novel memory-efficient event-based representation for samples that display sudden, irreversible transitions over time. As little as three image volumes suffice for a broad range of applications: the initial attenuations, the final attenuations and the \textit{local} transition times.
This third volume represents spatially distributed events on a continuous timescale instead of the discrete global time-frames.
We propose a method to iteratively reconstruct the transition times and the attenuation volumes. The dynamic reconstruction technique was validated on synthetic ground truth data and experimental data, and was found to effectively pinpoint the transition times in the synthetic dataset with a time resolution corresponding to less than a tenth of the amount of projections required to reconstruct traditional $\mu$CT time-frames.
\end{abstract}


\section{Introduction}

4D X-ray imaging is a powerful tool to observe ongoing dynamic processes inside optically opaque materials non-destructively. Hardware developments have led to the ability of performing tomoscopy, which is uninterrupted, continuous $\mu$CT acquisition at high frame rates while dynamic processes occur in-situ within the sample~\cite{garcia2021tomoscopy,dierick2014recent}.
The dynamic $\mu$CT data is typically acquired and processed in software as an image sequence of 3D attenuation volumes.
From the evolution of X-ray attenuation coefficients in the sequence of $\mu$CT images, the experimenter is able to perceive local changes in structure that occur in-between the consecutive scans.
As a result, in-situ $\mu$CT scanning methodologies have helped visualizing and studying dynamic processes for applications in various fundamental research domains~\cite{zwanenburg2021review}.
Prominent examples are fluid flow scans in porous media~\cite{bultreys2016imaging}, mechanical loading~\cite{jailin2017situ}, medical examination~\cite{chen2009temporal}, additive manufacturing~\cite{leung2018situ}, and pharmaceutical 
processes~\cite{moazami2022development}.
As a result, this quantitative visualization technique enables academia and industry to design more efficient processes or to understand failure mechanisms in the search for more durable development. \\

When dynamic changes occur in the sample during the acquisition, this acquisition becomes a methodological resource to observe the dynamics at the highest temporal resolution.
Two major challenges for dynamic $\mu$CT imaging are achieving sufficiently high temporal resolution and maintaining an overview over the ever larger datasets.
The first challenge stems from the observation that frame-based reconstructions are limited in their temporal resolution by the duration of the time-frames in which sufficient projections are acquired for 3D reconstructions. This duration increases at better spatial resolution and signal to noise ratio.
This is mainly an issue in applications with a relatively high dynamic speed compared to the current time scales that are available in lab-CT setups or at synchrotron facilities. 
For instance, fluid flow in porous media can display rapid pore-filling events with individual timescales around the millisecond~\cite{berg2013realtime}.
This is orders of magnitude below reported time-frame durations for continuous $\mu$CT scanning, which are around seconds for lab-CT and fractions of seconds for synchrotron imaging, considering the addition of peripheral equipment for in-situ control over the fluid flow~\cite{bultreys2016fast,pak2023design}. 
Unresolved dynamics that occur during the acquisition lead to motion artefacts, and are therefore often seen as a nuisance that should be mitigated in order to preserve an optimal image quality.
Therefore, more advanced reconstruction techniques are required to extract that valuable information from in-situ dynamic scans.
When higher temporal resolution is desired, radiographic techniques are often used to observe dynamics with temporal precision down to the projection level, albeit in a 2D projective view~\cite{armstrong2014subsecond}.
For this reason, many techniques were developed to bridge this gap between radioscopy and tomoscopy by modelling temporal changes in the sample at higher frequencies than the conventional rate of filtered backprojection methods.
These are mostly iterative methods instead of analytic reconstruction techniques, since they are more flexible to drop the assumption of a static sample and to exploit prior knowledge in spatial or temporal domain about the sample~\cite{myers2011dynamic}.
Two major complementary domains in iterative reconstruction for dynamic $\mu$CT are motion-compensated reconstruction~\cite{odstrcil2019initio,de2018motion,burger2018variational,zang2018spacetime,li2005motion,rit2009comparison} and sparse view or
limited view reconstruction techniques~\cite{goethals2022dynamic,arridge2022joint,nikitin2019four,mory2016motion,mohan2015timbir,kazantsev2015reconstruction,chen2009temporal,reed2021dynamic}.
These current state-of-the-art techniques for dynamic CT reconstruction result in 4D datasets that represent a time sequence of 3D frames.
Their gains in temporal resolution are therefore usually quantified by the decrease in number of projections while maintaining sufficient image quality in the spatial domain. \\

While the frame-based approach allows great dimensional freedom in the many individual frames of the limited view reconstructions, this reflects in a large memory consumption in the output of these methods.
This leads to the second challenge, namely processing large datasets while remaining sensitive to smaller details.
Often, data acquisition, reconstruction, segmentation and interpretation of dynamics are treated frame by frame~\cite{zwanenburg2021review}. 
This divide-and-conquer strategy inhibits a data processing pipeline that can ultimately trace back the derived dynamics to the original acquisitions, which is needed to confirm new experimental findings that are at the limits of the achievable temporal resolution. \\
To find a way around the large amount of data that is encountered in frame-based reconstruction techniques for dynamic $\mu$CT, our proposed approach is to impose a low-parameter-count model of the attenuation evolution inside individual voxels and to estimate these parameters \emph{directly} from the raw projection data, i.e. without reconstructing a sequence of many discrete CT volumes as a restrictive step.
The rationale behind this low parameter count is that many experiments are very sparse in terms of dynamics that occur, which makes it highly applicable.
Often, researchers are interested in the emergence of cracks~\cite{zhai2019high,jailin2017situ}, dissolution~\cite{moazami2022development} or precipitation~\cite{lee2023machine} of material, or displacement of fluids~\cite{bultreys2016imaging,myers2011dynamic,van2015iterative}.
In a fixed coordinate system, these dynamics can be presented as a set of events: localised transitions from the initial attenuation to the final attenuation of a selected region.
An additional advantage of using low-rank temporal models is that projection preprocessing techniques can be used with lower computation time to enable real-time reconstruction~\cite{mori2007candidate}.
One example of an efficient local event description, (i.e., for each voxel) is SIRT-PWC~\cite{van2015iterative}, which iterates between region-based SIRT reconstruction steps and an estimation of piecewise-constant (PWC) functions based upon the temporal sequence of reconstructed frames.
Similarly, Gao et al. used this to improve the temporal resolution for nanometre-scale dynamic imaging down to 12 minutes per time-frame~\cite{gao2024dynamic}.
To do this, they show that high-quality global time frames can be reconstructed from only 25 angles, using a piecewise-constant parametrisation.
This illustrates the strength of the event-based approach. \\

The proposed DYRECT technique demonstrates that it is possible to bypass the reconstruction of frames and to improve the temporal resolution directly to that of the projection-level, typically one to three orders of magnitude faster than the time scale of global time-frames.
Recently, Gorenkov et al.~\cite{gorenkov2024projection} demonstrated a projection-based preprocessing technique to achieve higher temporal resolution in fluid flow $\mu$CT scanning. 
The timestamps of local fluid displacements were pinpointed by subtracting the static content from the projections. These local indications were matched afterwards in the reconstructed frames by their similarity to motion-blurred regions.
To achieve this in a data-consistent manner, we bypass this separate matching procedure completely by integrating this event description in iterative reconstruction: the transition times are stored directly in the sample volume, which indicates when there is a change in the attenuation of the voxel.
This builds upon the reconstruction concept of flexible spatio-temporal decompositions, where low-rank temporal basis functions are used to express the evolution of the attenuation~\cite{boigne2022towards,nikitin2019four,iskender2024redpsm}, allowing changes at other projection times than the preset coarse time steps.
However, in case that the sample exhibits multiple events at different times, that approach would require an increasing number of global basis functions and can still be affected by motion artefacts due to the scan time per (limited) projection window.
Therefore, changes in the sample should be represented (1) by local time-related event parameters such as transition times (cf.~\cite{van2015iterative,gao2024dynamic}), \textit{and} (2) starting from individual projections (cf.~\cite{gorenkov2024projection,boigne2022towards,nikitin2019four,iskender2024redpsm}) to get the maximal temporal resolution.\\

This work shows the first results of dynamic reconstruction of events on a continuous projection-level timescale. 
Section~\ref{ssec:meth:rec} covers the steps taken to estimate the transition times from the projection data in an iterative approach. The method was evaluated on experimental data in section~\ref{ssec:res:coalescence}, and realistically simulated data, introduced in section~\ref{ssec:res:simulated}. 
Since the local event-based representation had not yet been used with temporal resolution down to the continuous projection level, this opens new questions regarding the limits on temporal resolution. 
Therefore, we discuss the interplay of the acquisition angle and the temporal resolution in sections~\ref{ssec:res:angular_structure} and ~\ref{ssec:res:angular_flow}.\\




\section{Methods}

\subsection{Event-based CT reconstruction (DYRECT)}
\label{ssec:meth:rec}

\begin{figure}[!ht]
    \centering
    \includegraphics[width = \columnwidth]{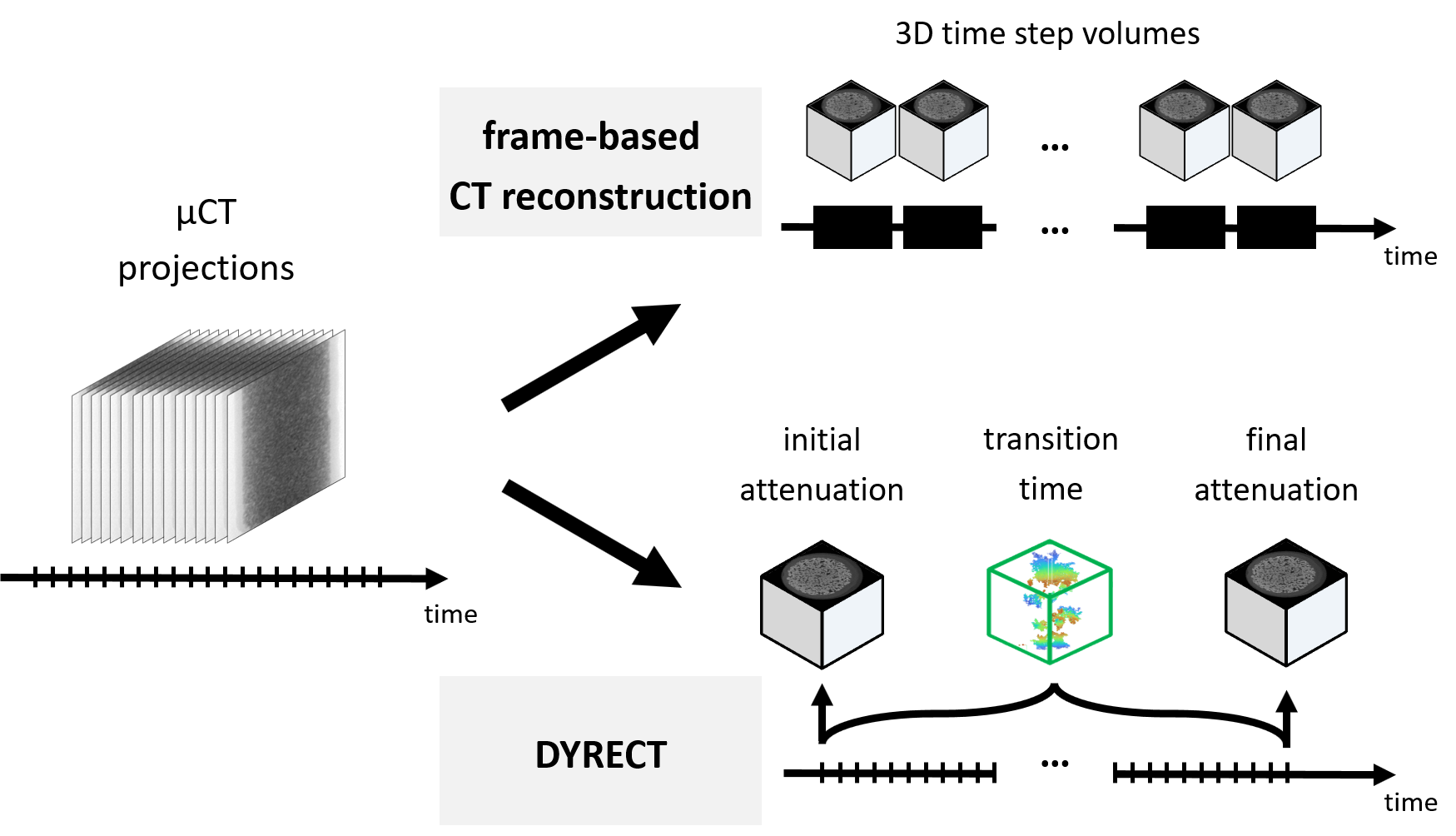}
    \caption{Schematic comparison between the proposed event-based reconstruction (DYRECT) and conventional frame-based reconstruction techniques. DYRECT describes events with fine temporal resolution at the projection level, using a single volume of local transition times. This mitigates temporal blurring associated with the longer frame sequence of coarse-resolution time step volumes.}
    \label{fig:schematic}
\end{figure}

We use a 4D $\mu$CT reconstruction technique, combining all projections in the scan, to compute the event parameters. As is the case for motion-compensated reconstruction techniques, this paradigm requires the implementation of a ray-tracer that incorporates the temporal dependency of the acquired projections.
As illustrated in figure~\ref{fig:schematic}, we choose to represent the 4D dynamic volume by a small set of local parameters, stored in Cartesian volumes, that describe the evolution of each voxel over time $t$.
Despite its simplicity, we consider the \textit{single step model} to be suited well in two broad categories of irreversible dynamics:
\begin{enumerate}
    \item material precipitation~\cite{lee2023machine} or removal~\cite{moazami2022development,vanbillemont20204d};
    \item propagation of homogeneous bulk materials and fluids~\cite{bultreys2016imaging,myers2011dynamic} and crack emergence~\cite{zhai2019high,jailin2017situ}.\\
\end{enumerate}

The prime applications of the event parametrization are those with stationary features in attenuation transition. While the single transition model makes a strong assumption with only two global sample states explicitly in memory, there can be an arbitrary number of global sample states at times in-between with configurations where some regions have transitioned and some not. Additionally, the transition is assumed to be instantaneous. If the transition takes longer, because of slower concentration changes~\cite{vanoffenwert2021fast} or partial volume effects, a gradual transition model like a sigmoid function should be used.
The second category of applications are those with visible motion, like materials that crack~\cite{jailin2017situ} or fluids that propagate through a static porous medium~\cite{bultreys2016imaging}.
This resolves the challenge that explicit local motion models have, e.g. using deformation vector fields (DVFs), since they are designed for smooth motion, and cannot capture sudden accelerations. Discontinuous dynamics are therefore better captured in the event-based framework, particularly near boundaries of materials. \\

In this single step model, we consider the evolution within the voxel with index $j$ to be fully determined by three parameters: the voxel's initial attenuation $\mu_{A,j}$, its final attenuation $\mu_{B,j}$ and its transition time $t_j^*$.
\begin{equation}
    \mu_{\mathrm{step}}(t; \mu_{A,j}, \mu_{B,j}, t_j^*) = 
    \mu_{A,j}~\mathcal{H}(t_j^* - t) + \mu_{B,j}~\mathcal{H}(t -  t_j^*)
    \label{eq:local_decomposition}
\end{equation}
The Heaviside function $\mathcal{H}$ models a discrete step from the initial phase to the final phase.
Increasing $t_j^*$ leads to a longer temporal support on the attenuation $\mu_{A,j}$. Conversely, decreasing $t_j^*$ means that the final  phase $\mu_{B,j}$ initiates earlier. \\

Our proposed approach is to estimate the event parameters in an iterative way to obtain a data-consistent representation of the sample's evolution.
The method relates to the static case update schedule SIRT (Simultaneous Iterative Reconstruction Technique), where the attenuation coefficients of the $N_V$ voxels in the sample are updated iteratively by cycles of 1) forward projection, 2) calculation of correction terms and 3) backprojection of those correction terms. 
In algebraic notation, the new estimate of the static volume $\bm{\mu} \in \mathbb{R}^{N_V}$ at iteration $it$ over all $N_T$ projections in the scan is given by~\cite{gregor2008computational}:
\begin{equation}
    \bm{\mu}^{(it)} = \underbrace{\bm{\mu}^{(it-1)} + C A^T \underbrace{R \left(\mathbf{p} - \underbrace{A \bm{\mu}^{(it-1)}}_{1} \right)}_{2}}_{3}. \\
    \label{eq:sirt_matrix}
\end{equation}

The projection values are given by the optical depth $p = -\ln{(\sfrac{I}{I_{0}})}$ for the attenuation of X-rays, corresponding to Lambert-Beer's attenuation law.
$A$ is the sparse system matrix $\in \mathbb{R}^{N_T N_P \times N_V}$ that gives the intersection lengths for the X-rays towards pixels $i$ through voxels $j$. $R$ and $C$ are the diagonal matrices of inverse row and column sums of the system matrix. The $i$'th diagonal element of R, $r_{ii} = 1/ \sum_j^{N_V} a_{ij}$, corresponds to the inverse of the ray length towards one pixel $i$ recorded at a given projection time $t(i)$. There are $N_P N_T$ projected pixels in total, where $N_P$ is the number of physical pixels per single projection. \\

To obtain local estimates of the dynamic events from a $\mu$CT scan, the projection and backprojection steps of the SIRT technique are altered to incorporate the temporal dependency of the projection acquisition. \\

\paragraph{Local event-based $\mu$CT projections}

Denoting 3D time slices of the 4D volume by $\bm{\mu}_t$, where $t$ is the projection index, 
the static-volume system matrix $A$ is replaced by the dynamic-volume system matrix $\tilde{A}$ comprising single-projection system matrices $A_t$ $\in \mathbb{R}^{N_P \times N_V}$ for the corresponding projection vectors $\mathbf{p}_t$ at every time index $t \in [0, 1, 2, \dots, N_T - 1]$. 
The forward operator for the projection estimate $\hat{\mathbf{p}}$ can be formally written using a block diagonal expression:

\begin{align}
\begin{bmatrix}
\hat{\mathbf{p}}_0 \\  
\hat{\mathbf{p}}_1 \\
\vdots \\
\hat{\mathbf{p}}_{N_T-1} \\
\end{bmatrix}
&=
\underbrace{
\begin{bmatrix}
A_0  & \mathbf{0} & \cdots      & \mathbf{0}         \\
\mathbf{0}  & A_1 & \cdots &  \mathbf{0}\\
\vdots & \vdots  & \ddots & \vdots \\
\mathbf{0} & \mathbf{0} & \cdots & A_{N_T-1}
\end{bmatrix} 
}_{\tilde{A}}
\underbrace{
\begin{bmatrix}
\bm{\mu}_0 \\  
\bm{\mu}_1 \\
\vdots \\
\bm{\mu}_{N_T-1} \\
\end{bmatrix}
}_{\tilde{\bm{\mu}}}
\label{eq:projection}
\\
\hat{\mathbf{p}}_t &\stackrel{(\ref{eq:local_decomposition})}{=}
\begin{aligned}[t]
    & A_t (\bm{\mu}_A \odot \mathcal{H}(\mathbf{t^*} - t)) \\
    + & A_t (\bm{\mu}_B \odot \mathcal{H}(t - \mathbf{t^*}))
    \label{eq:step_projection}\\
\end{aligned}
\end{align}

$\odot$ denotes elementwise multiplication of the attenuation coefficients and the time-dependent weights in the volume domain.\footnote{Observe that if $\mathbf{t}^*$ were a global scalar for the volume, these weights could be included after performing static projections:
\begin{equation}
    \hat{\mathbf{p}} = (A \bm{\mu}_0) \odot \mathcal{H}(t^* - \mathbf{t}_p)
    + (A \bm{\mu}_{N_T-1}) \odot \mathcal{H}(\mathbf{t}_p - t^*)
\end{equation}
Where $\mathbf{t}_p$ is the vector of acquisition timestamps for each of the $N_P N_T$ pixel records. This resembles a formulation of global breakpoints~\cite{boigne2022towards}.}
The compact representation using a local transition map requires that the temporal weight is multiplied during ray tracing. 
Therefore, all three parameter volumes ($\bm{\mu}_A, \bm{\mu}_B, \mathbf{t^*}$) are used during the forward projection on the GPU.
At each evaluation of the line integral, the time-dependent attenuation is calculated locally from these volumes.
This addresses the first part of the reconstruction, namely the dynamic projection simulation. \\

\paragraph{Local event estimation in $\mu$CT backprojection}


The 4D attenuation volume $\tilde{\bm{\mu}}$, in this case parametrized by three 3D volumes, is updated in the back-projection step. 
Since this update affects the complete time profile of a given voxel with index $j$, we construct the time-dependent correction terms $\delta_j(t)$ by sampling the corresponding pixel positions in the sinogram domain with bilinear pixel interpolation.
This corresponds to:
\begin{equation}
    \delta_j(t) = C_{t,j} A_t^T R_t \left(\mathbf{p}_t - \hat{\mathbf{p}}_t \right)
    \label{eq:correction_terms}
\end{equation}
This time-dependence emerges as the pixels from different projections corresponding to a selected voxel $j$ are recorded at a different known acquisition time, indexed by $t$. $C_{t,j}$ is row $j$ of the single-projection diagonal matrix $C_t$.
Note that the backprojection for the static case in equation~\ref{eq:sirt_matrix} immediately reduces the correction signal to a single average correction term per voxel\footnote{There is approximately one pixel per 2D projection image at time $t$ that corresponds to a selected voxel $j$.
Therefore, the elements of the matrix $[CA^T]_{ji} = \sfrac{a_{ij}}{\sum_i^{N_P N_T}{a_{ij}}}$, which are used to compute a weighted sum of the correction terms $i$ for voxel $j$, can be approximated by a $\sfrac{1}{N_j}$ for corresponding pixels~\cite{de2017fast}.
$N_j$ is here calculated as the number of 2D projections where the voxel $j$ is in the field of view.}, thus ignoring any time-dependence.
As the mean correction term in the static case reflects a least-squares minimization, the back-projection step in the time-dependent case can also be conceived as performing a direct least-squares fit per voxel of the parametrized evolution function to the projection-corrected version of the previous estimate. 
In general terms denoted by an event parametrization $\bm{\varepsilon}_j = [\bm{\mu}_A, \bm{\mu}_B, \mathbf{t}^*]_j$ per voxel $j$ (cf. equation~\ref{eq:local_decomposition}), we would get the following update equation for iteration $it$: 
\begin{equation}
    \bm{\varepsilon}_j^{(it)} = \underset{\bm{\varepsilon}_j}{\arg \min} \sum_t^{N_T}\left[\mu_{\mathrm{step}}(t; \bm{\varepsilon}_j^{(it)}) - \mu_{\mathrm{step}}(t ; \bm{\varepsilon}_j^{(it-1)}) - \delta_j(t) \right]^2
    \label{eq:update_optim}
\end{equation}

In practice, however, this update can make large changes to the transition time, strongly affecting the transition time of nearby voxels.
This makes that the final estimate of the transition time can be unstable, with oscillations depending on the iteration number.
Therefore, we adapt the optimisation of the event parameters by first updating the transition time $t^*_j$ based on imbalance of correction terms $\delta_j(t)$ at projection times before $(t_{A,j})$ and after $(t_{B,j})$ the current transition time $t^{*(it-1)}_j$.
The intuition behind this is that correction terms show the strongest discrepancies in the temporal proximity of the currently estimated transition time, if it is still incorrect.
When only the transition time $t^*$ is sought in the reconstruction, the difference in mean values could be used. However, when $\mu_A$ and $\mu_B$ are not held fixed, a third independent value must be estimated from the sequence of correction terms.
The array of $N_j$ correction terms $c$ from equation~\ref{eq:correction_terms} is converted into two growth terms $\sigma_{A,j}$ and $\sigma_{B,j}$
over projection windows before and after the current estimated transition time, respectively.
The calculation of the growth was implemented as a single sampling pass over the correction terms with recursive updates to calculate the covariance between the correction terms and the projection time~\cite{socha2017optimization} (without dividing by the variance over the time domain). 
The growth values were calculated over symmetric windows of $360\degree$ before ($N_{A,j}$ projections with voxel $j$ in the field of view) and after ($N_{B,j}$ projections) the prior estimated transition time, which mitigates the influence of structural biases by exploiting their repeating appearance in the circular scanning trajectory.
Currently, this restricts the method to analyse scans that span at least $3$ full rotations with $t_{360\degree}$ projections in a circular trajectory. \\

{
\smaller[0.5]
\allowdisplaybreaks
\begin{align}
    &\sigma_{A,j} = \dfrac{1}{N_{A,j}} \sum_{t_{A,j}}^{N_{A,j}} \left(t_{A,j} - \overline{t_{A,j}}\right) \left(\delta_j(t_{A,j}) - \overline{\delta_j(t_{A,j})}\right)
    \nonumber \\
    & \quad \text{with $t_{A,j} \in S_{A,j}^\mathrm{(it)} :=\left[t^{*(it-1)}_j - t_{360\degree}, t^{*(it-1)}_j\right]$,} \label{eq:growth_before}\\ 
    &\sigma_{B,j} = \dfrac{1}{N_{B,j}} \sum_{t_{B,j}}^{N_{B,j}} \left(t_{B,j} - \overline{t_{B,j}}\right) \left(\delta_j(t_{B,j}) - \overline{\delta_j(t_{B,j})}\right)
    \nonumber \\
    & \quad \text{with $t_{B,j} \in S_{B,j}^\mathrm{(it)} := \left[t^{*(it-1)}_j, t^{*(it-1)}_j + t_{360\degree}\right]$, } \label{eq:growth_after}
\end{align}
\larger[0.5]
}

\begin{figure}[!ht]
    \centering
    \includegraphics[width = \linewidth]{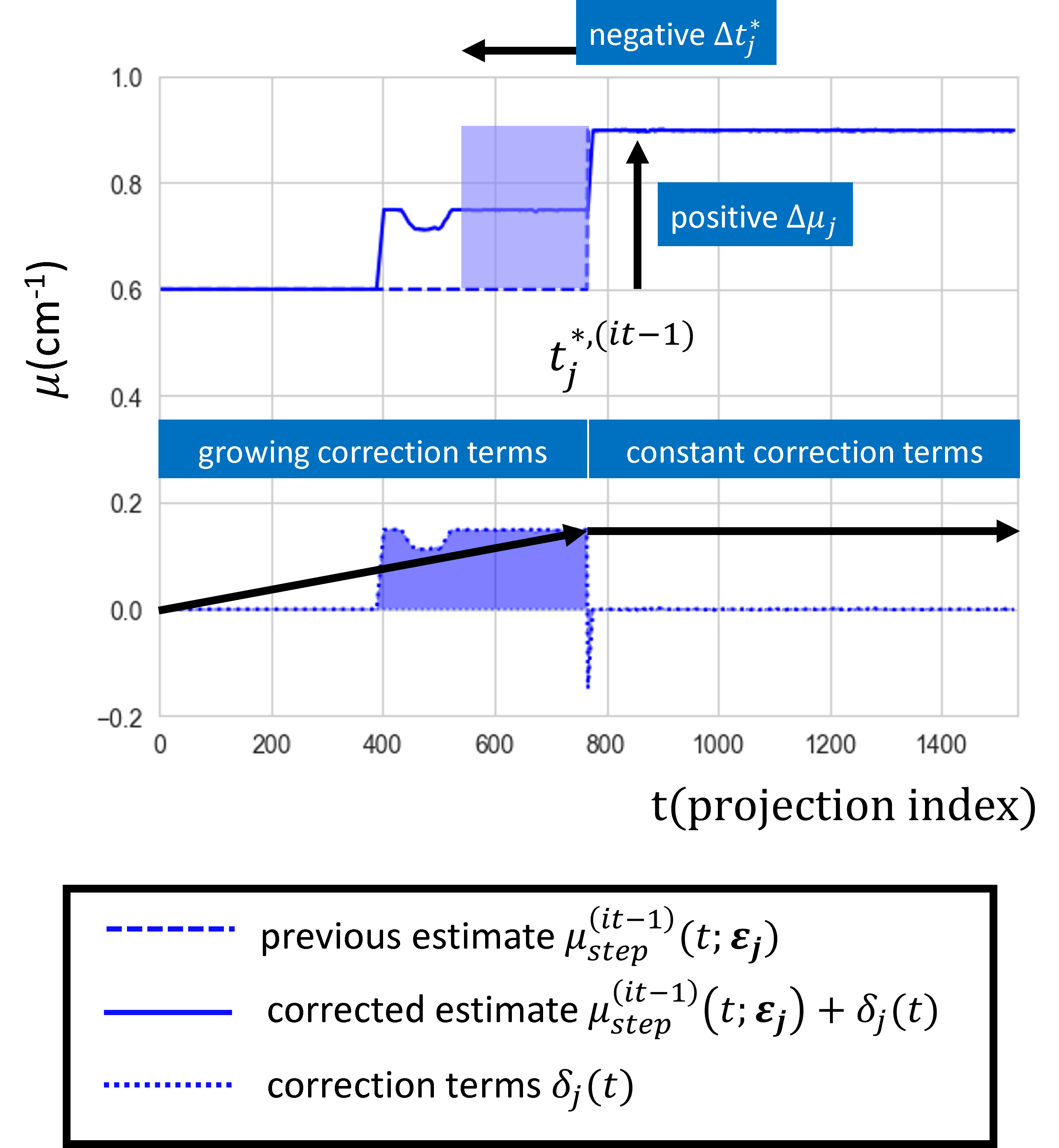}
    \caption{Illustrative update of the transition time $t_j^*$ of a single voxel $j$. The dashed blue line indicates the prior estimate $\mu_j(t)$ of the voxel, based on the three parameters $t_j^{*(it-1)} = 768$, $\mu_{A,j}^{(it-1)} = 0.6\,\mathrm{cm}^{-1}$, and $\mu_{B,j}^{(it-1)}= 0.9\,\mathrm{cm}^{-1}$. The full line is the virtual corrected attenuation curve $\mu_\mathrm{step}(t;\varepsilon_j)+\delta_j(t)$, by addition of the correction terms below, shown as the dotted line. The blue area at the bottom, between projection index $t = 400$ (ground truth) and $768$ (current estimate) indicates that the attenuation for those projections should be increased. Since the projections between $t = 0$ and $400$ have zero correction terms, this growing trend indicates that \textit{not} $\mu_{A,j}$ should be updated, but that the final higher attenuation phase should start earlier by shifting $t_j^*$ to a lower projection time.}
    \label{fig:balance_tstep}
\end{figure}

The overline symbol denotes the arithmetic mean over the selected projection indices ($S$) in the rotation before or after the current transition time $t_j^{*(it-1)}$.
The update of the transition $t_j^*$ for one selected voxel $j$ is illustrated in figure~\ref{fig:balance_tstep}. To find the optimal transition time for the selected voxel, the update minimises the difference in these growth terms before and after the prior estimate $t_j^{*(it-1)}$ of the event.
The change $\Delta t_j^*$ in the update of $t_{j}^{*}$ is modified based on the prior difference $\Delta \mu_j$ in attenuation values.
In equation~\ref{eq:change_event}, this reflects the lower confidence put in voxels with a small difference between the initial and final phase, while simultaneously converting the dimension of the growth terms $\sigma_{A,j}, \sigma_{B,j}$ to a dimension of time. 
$\lambda_\Delta = 0.1\,\mathrm{cm}^{-1}$ normalizes the difference values to the user-defined maximum value. 
To avoid division by zero, a small value $\epsilon = 10^{-5}$ is added in the denominator with equal sign as $\Delta \mu_j$.
Furthermore, larger updates are restricted to the time $t_{180\degree}$ corresponding to a half rotation in equation~\ref{eq:limit_change_event} and  relaxed by the factor $\lambda_t = 0.6$ ($0<\lambda_t<1$).
\\

{
\allowdisplaybreaks
\begin{align}
    & \Delta \mu_j = \mu_{B,j} - \mu_{A,j}
    \label{eq:attenuation_difference}\\
    & \Delta t_j^* = \left(\sigma_{B,j} - \sigma_{A,j} \right)\dfrac{\min(|\Delta \mu_j| / \lambda_\Delta, 1)}{\Delta \mu_j + \mathrm{sign}(\Delta \mu_j) \cdot \epsilon}
    \label{eq:change_event}\\
    & \Delta t_j^* \leftarrow \mathrm{clip}\left(\Delta t_j^*, -t_{180\degree}, + t_{180\degree} \right)
    \label{eq:limit_change_event}\\
    &t_{j}^{*(it)} = t_{j}^{*(it-1)} + \lambda_t~\Delta t_j^*
    \label{eq:update_event}
\end{align}
}

After updating the transition time $t_j^*$, the correction terms $\delta_j(t)$ are sampled a second time for the targeted voxel $j$ to estimate the attenuation before and after the updated transition time. The corrected attenuation values are averaged in equations~\ref{eq:attenuation_before} and~\ref{eq:attenuation_after}. In the final updates, the values are relaxed by respective factors $\lambda_A$ and $\lambda_B = 0.8$ to achieve a more gradual update and improve stability.

{
\smaller[0]
\allowdisplaybreaks
\begin{align}
    &\mu^{(it)}_{A,j} = \mu^{(it-1)}_{A,j} + \lambda_A \left(\overline{\mu_\mathrm{step}(t_{A,j}; \varepsilon_j^{(it-1)}) + \delta_j(t_{A,j})} - \mu^{(it-1)}_{A,j} \right)
    \label{eq:attenuation_before} \\
    &\mu^{(it)}_{B,j} = \mu^{(it-1)}_{B,j} + \lambda_B \left(\overline{\mu_\mathrm{step}(t_{B,j}; \varepsilon_j^{(it-1)}) + \delta_j(t_{B,j})} - \mu^{(it-1)}_{B,j} \right)
    \label{eq:attenuation_after}
\end{align}
}


\paragraph{Update sequence: Ordered subset of projections}

Each projection and backprojection step is calculated based on ordered subsets $S$ of the acquisition sequence~\cite{beister2012iterative}. 
This speeds up equations~\ref{eq:growth_before} and~\ref{eq:growth_after}, thus improving memory efficiency and convergence speed~\cite{xu2010efficiency}.
The full set of projections is split in separate subsets of approximately equal size, and each subset contains projections that are randomly selected from the full set within the user-defined projection window.
The projection window corresponds to the acquisition time of the first projection until the last projection. Therefore, it should be chosen such that the sample dynamics within the projection window can be described accurately by the model for the optimal set of parameters. 
This means that the size of the subsets is linked to the convergence speed and stability of the algorithm. Algorithms using smaller subsets, such as SART~\cite{beister2012iterative}, employ a more greedy update step and are less stable. Algorithms using larger subsets, such as SIRT, require more iterations to steadily converge towards the solution.
For dynamic CT reconstruction, there is an additional reason to use larger subsets. As in the first step of the backprojection, the transition time $t_j^*$ is updated independently in each voxel based upon the growth terms. The computation of these terms is less stable when a small selection of projections is sampled in the backprojection. The update of $t_j^*$ affects the temporal support of the initial and final attenuation value changes. Therefore, each subset should offer a broad temporal range of projections to guarantee the stable reconstruction. \\
\\

\paragraph{Weighted backprojection}
To further improve image quality, additional priors can be incorporated in the reconstruction.
One optional prior is the indication of static and dynamic regions in the volume. As indicated above, these can be used to subtract the static signal in tomographic reconstruction~\cite{deurydice2018dynamic}, or to suppress the variability of the static regions~\cite{myers2011dynamic,van2015iterative}.
To implement this, a volume of weights $\mathbf{w}$ is defined that indicates which regions are more likely to exhibit dynamic changes throughout the scan~\cite{heyndrickx2020improving}. Higher weights are attributed to these regions in the backprojection step, encouraging sharper updates in those regions. This replaces equations~\ref{eq:projection},~\ref{eq:attenuation_before}, and~\ref{eq:attenuation_after} to incorporate weights. 
The concept of weighted back-projection was used for the reconstruction of the simulation study of fluid flow in a porous sandstone. 
The selection of the weights was modified in this work to conceive a hybrid estimation of static and dynamic regions on the fly.
This is based on the difference between the initial and the final attenuation of the voxel, which is computed during the reconstruction. 
\\

\subsection{Ground truth dataset: dynamic CT simulation of fluid flow in a porous medium}
\label{ssec:meth:sim}
In the simulation experiment designed to validate the reconstruction method, the phantom mimicked multiphase flow in a porous sandstone sample. 
This kind of flow often exhibits single attenuation transitions per position while one fluid is pumped through the sample to displace the other.
Within the static matrix (i.c. sandstone), the structure of the porous network impacts the various characteristics of the fluid flow on a local scale~\cite{berg2013realtime}.
Different flow patterns were simulated in the single 4D simulation, by selection of the transition times. The flow direction of the fluid meniscus is given by the 3D spatial gradient unit vector of the transition time. Some flows were oriented predominantly horizontally, while others were directed vertically. It was expected that the vertical flows, along the rotation axis, were easier to distinguish since this can also be studied using radioscopy (time-resolved radiography). The speed of the flow relates to the gradient magnitude of the transition times $|\nabla_\mathbf{x} t^*(\mathbf{x})|$. By controlling this, there was a variety of smooth and sudden flows. The sole restriction was that the transition times changed monotonically throughout the spatial domain, to achieve a contiguous fluid front. \\

The synthetic phantom and scan conditions were based on a real scan and reconstruction of a sandstone sample. The number of voxels and pixels in the horizontal direction was equal to 989, and the detector had 1528 rows. Per 360\textdegree\,rotation, 1482 cone beam projections were simulated.
The dynamics were simulated in the second rotation of three in total.
The simulated projections and reconstructions are shared in the online repository \url{doi.org/10.17605/OSF.IO/64ERH}. \\


\subsection{Experimental dataset: bubble coalescence}
\label{ssec:meth:exp}

We wish to evaluate whether DYRECT leads to an appreciable improvement in time resolution. To test this, the technique was applied on a given experimental dataset captured in challenging conditions~\cite{garcia2021tomoscopy}.
The experiment was designed to scan bubble coalescence in liquid metallic foam, meaning that two neighbouring bubbles in the metallic foam became one by rupture of the separating film in between the two. 
The goal of our study was to capture this coalescence event in the reconstruction at higher temporal resolution.
The used selection of the scan consisted of three consecutive complete revolutions, with the bubble coalescence event during the second.
The sample was a metallic foam of AlSi8Mg4 with an inner diameter of 1.2\,mm, contained in a small boron nitride crucible. By in-situ laser heating during the CT experiment, the sample underwent stages of bubble nucleation, inflation and coalescence. 
This was captured at high CT scanning rates with 500 full rotations per second, acquiring 80 projections per 360\textdegree\,rotation with a pixel size of 2.75\,$\mu$m at the TOMCAT beamline of the Swiss Light Source synchrotron.
Each projection therefore corresponds to a time interval of 25\,$\mu$s. \\

\section{Results}
\label{sec:results}

\subsection{Reconstruction of bubble coalescence experiment: consistent with difference sinograms}
\label{ssec:res:coalescence}

\begin{figure}[!ht]
    \centering
    \includegraphics[width = 0.8\linewidth]{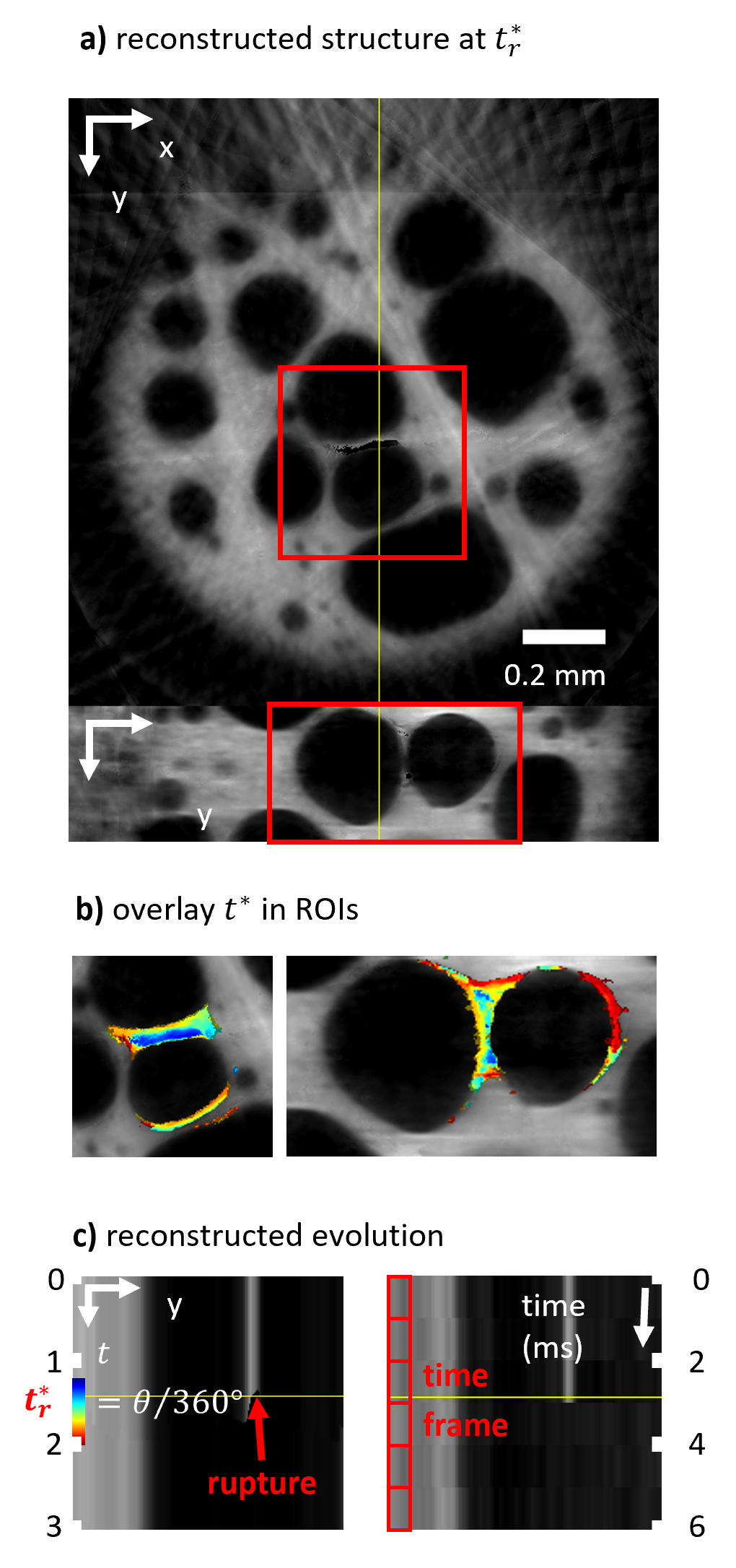}
    \caption{Horizontal and vertical cross-sections (\textbf{a}) of bubble coalescence experiment at the initial rupture moment, reconstructed using the DYRECT technique. The overlay images (\textbf{b}) show the transition times $t^*$ of the indicated regions of interest. The temporal cross-sections (\textbf{c}) of a DYRECT reconstruction were compared to those made with a SART reconstruction. The estimated time of rupture during the bubble coalescence is indicated by $t^*_r$ .
    }
    \label{fig:bubble_coalescence_rec}
\end{figure}

We used the DYRECT method to reconstruct the dynamic CT experiment of bubble coalescence in a metallic foam (section~\ref{ssec:meth:exp}).
The reconstruction technique made 10 iterations over the 240 projections, covering three $\mu$CT rotations of scanning the dynamic process at an acquisition rate of 500 rotations per second.
Even at those high scanning rates, film rupture is one type of dynamics that is challenging to reconstruct without motion artefacts due to the fast speed relative to the CT acquisition rate.
From the three reconstructed event parameter volumes $\bm{\mu}_{A}$, $\bm{\mu}_B$ and $\bm{t}^*$, three 2D slices are displayed in figure~\ref{fig:bubble_coalescence_rec} spanning the four dimensions. The first two are spatial cross-sections in the horizontal ($xy$) and vertical ($yz$) direction (a) at the estimated time of rupture $t^*_r$, and the third is a temporal cross-section, perpendicular to the ruptured film wall. 
The temporal cross-section of the DYRECT reconstruction shows that the film rupture between the two neighbouring foam bubbles happened at the rupture time $t^*_r$ corresponding to the acquisition angle $\theta^*_r = 504\degree$, or $1.4$ rotations (2.8\,ms after the first projection). The film wall completely disappeared at the time corresponding to $1.75$ rotations, or 0.7\,ms later. To indicate the usual temporal resolution for parallel beam, a SART reconstruction was made for the corresponding 6 180\textdegree\,rotations for three iterations over the 40 projections per time step.
This indicates that the single step model is a reasonable approximation that enables to model the evolution of finer details like the thinning of the film wall.
In the selected regions of interest, the transition map was overlaid on top of the initial reconstruction on areas with most change towards the final state (b). This map indicates that the deformation at the farther side of the right bubble was delayed by 1.0\,ms, which corresponds to the time to cover half a rotation in that scan. \\

\begin{figure}[!ht]
    \centering
    \includegraphics[width = \linewidth]{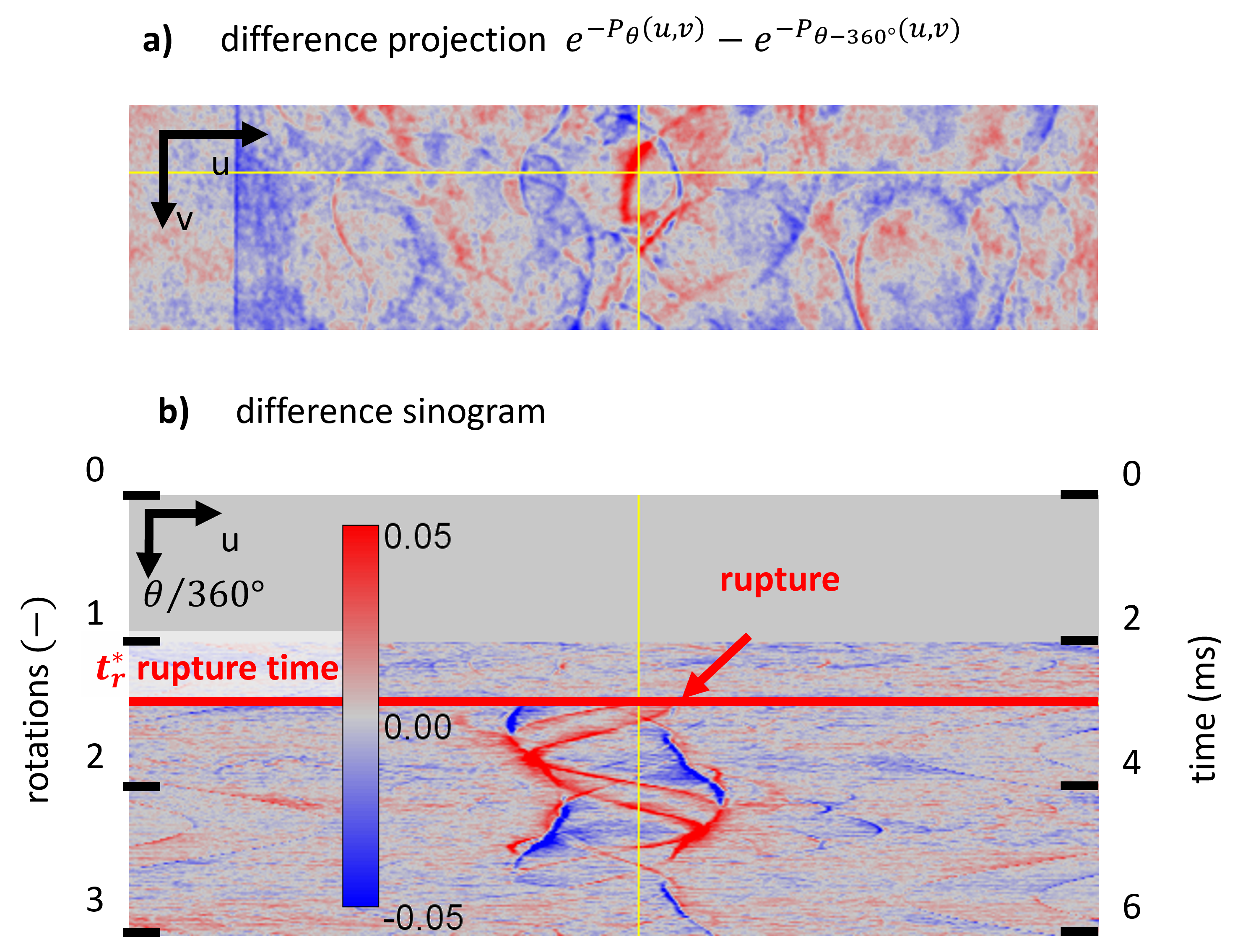}
    \caption{The reconstructed time of bubble coalescence $t^*_r$ corresponds to the time determined independently using the sinogram event detection technique. These difference projections (\textbf{c}) and sinograms (\textbf{d}) with the diverging colour map show the sample changes in the detector domain.
    }
    \label{fig:bubble_coalescence_proj}
\end{figure}

To verify the accuracy of the reconstructed transition times, a complementary technique was used to detect the events in the sinogram domain.
If it is the only event, the sinogram offers a good way to find the transition time.
To assert that the time resolution improvement is real, the DYRECT-pinpointed transition times needed to be consistent with changes observed independently in the sinogram slices (before and after the specific time).
This is indicated in figure~\ref{fig:bubble_coalescence_proj}. 
To focus on the dynamic content only, these images show the difference projection and sinogram with respect to the previous rotation of the continuous acquisition. This view ignores the static content of the acquisition and reveals what has changed with respect to the prior reference.
In the diverging colour map, the grey colour indicates that the normalised projected intensity remained constant, while red and blue indicate that the attenuation along the line from the source to the detector pixels decreased or increased, respectively.
The comparison between this difference sinogram and the DYRECT reconstruction (red horizontal line) shows that the start of the rupture event $t^*_r$ was estimated accurately with flexibility down to the projection level.
\\

\subsection{Angular dependency of temporal accuracy on direction of spatial structures and flow field}
\label{ssec:res:angular_structure}

\begin{figure}[!ht]
    \centering
    \includegraphics[width = 0.8\linewidth]{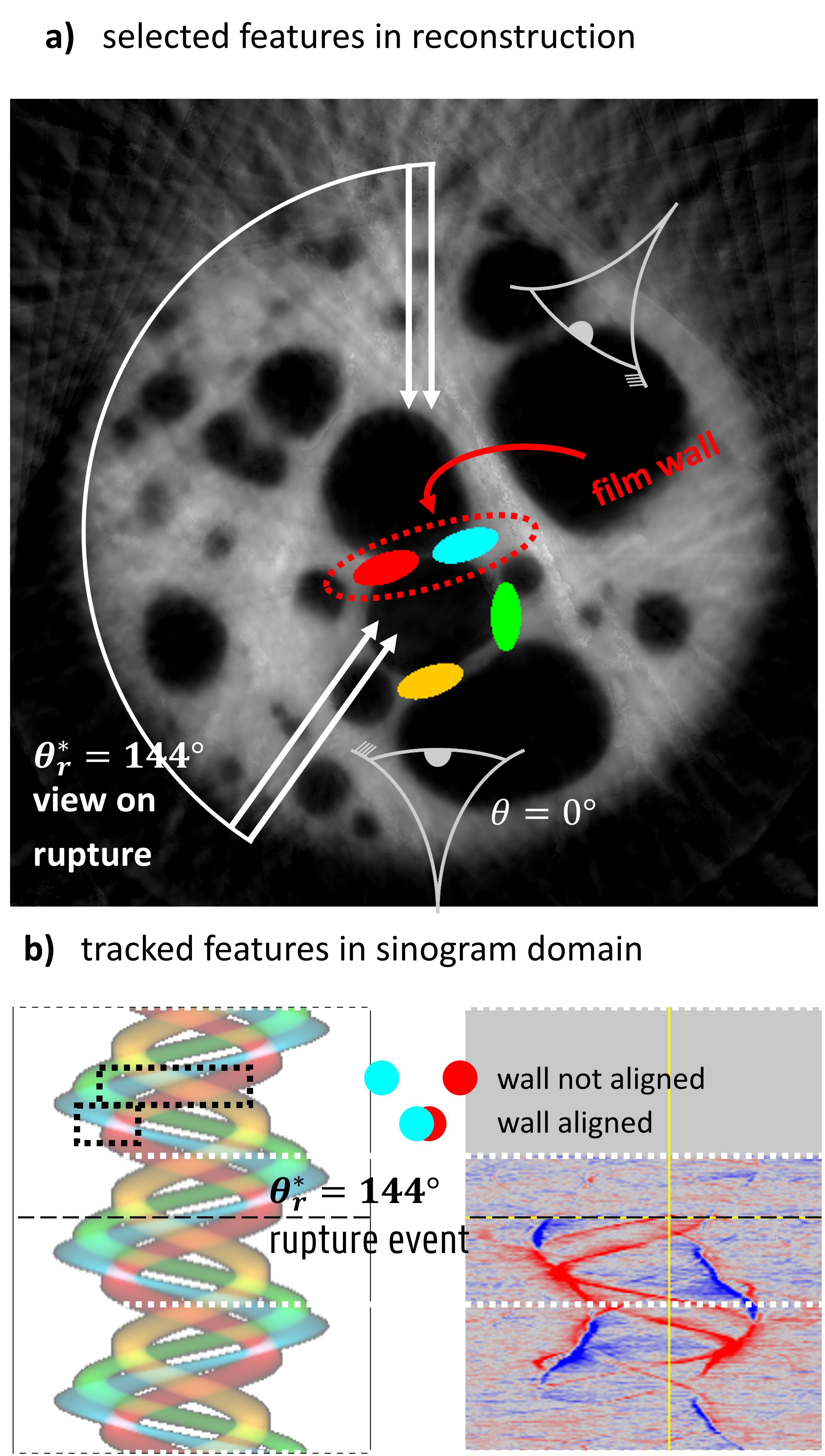}
    \caption{Tracking the selected features in the sinogram domain (red, cyan, green and yellow ellipses) reveals their evolution at a higher temporal resolution than possible with reconstruction techniques that define global time windows. The study indicated that the alignment of the dynamic internal features with respect to the optical axis may affect the accuracy. When the wall aligns with the optical axis, the (red and cyan) tracks in the sinogram domain overlap which causes ambiguity in the event localisation along the viewing direction.}
    \label{fig:bubble_tracking}
\end{figure}

To study the interplay between the orientation and the temporal resolution of the dynamic event detection, we compared the orientation of the reconstructed dynamic features to the acquisition angle at the instantaneous transition time, which is at the time scale of a single radiograph.
When two evolving structures align in the time-dependent acquisition geometry, it may affect the ability of the reconstruction algorithm to estimate the proper transition time.
In figure~\ref{fig:bubble_tracking}, different regions affected by the rupture event were tracked in the sinogram domain to gain intuition in the difference sinogram.
In the parallel beam geometry, the contrast of the film wall was highest at the acquisition angle of $\theta$ = 72\textdegree.
This does not align with the acquisition angle at the rupture angle at $\theta^*$ = 144\textdegree. 
Therefore, there was a distinct view on the evolution along the plane of the film wall, but to a smaller extent in the transverse direction. \\

It is conceivable that the physical evolution of the rupture event was not unambiguously reconstructed. 
Indeed, in figure~\ref{fig:bubble_coalescence_rec}, the film wall appears to become thinner starting from one side during the rupture, which conflicts with the intuition that the thinning of this film is a symmetrical process.
This ambiguity is a fundamental limitation of the single-source CT acquisition process, where the result is directly affected by the interaction between the internal dynamics and the time-dependent geometry of the acquisition. \\
\\

\subsection{Reconstruction of simulated data: local temporal resolution gain}
\label{ssec:res:simulated}

\begin{figure}[!ht]
    \centering
    \includegraphics[width = 0.9\linewidth]{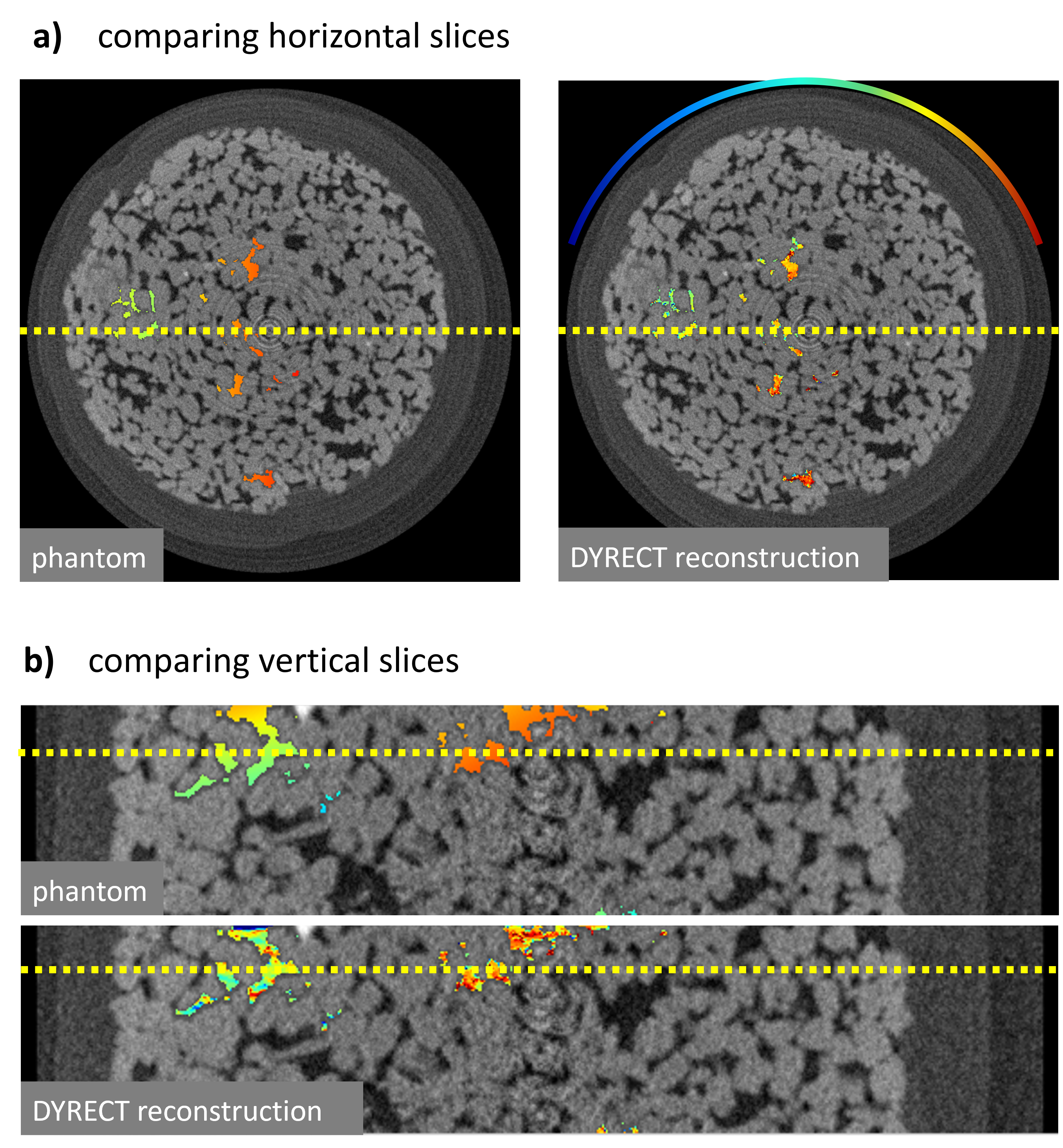}
    \caption{Comparison of the ground truth dynamic dataset and the outcome reconstructed using the proposed DYRECT method, visualized by 2D vertical (\textbf{a}) and horizontal (\textbf{b}) overlay slices of the coloured transition times in the dynamic regions and grey attenuation values in the static regions.
    The coloured arc in the reconstructed slice indicates the time-dependent viewing direction of the circular cone-beam CT system onto the dynamic process that is ongoing in the sandstone.
    }
    \label{fig:rec_simulation_overlay}
\end{figure}

The temporal resolution gain was evaluated using artificial ground-truth data. 
Figure~\ref{fig:rec_simulation_overlay} illustrates the outcome of the reconstruction method. 
Only the transition times were estimated in this reconstruction, while the start and final volumes remained static, initialised with exact ground-truth knowledge. This was done to study the transition time optimisation independently from the attenuation domain.
In experimental work, this is also a viable approach to scan the initial and final states under static conditions at higher image quality.
The grey values are the X-ray attenuation coefficients of the regions that remained static throughout the three CT rotations. The coloured overlay indicates the transition time of the dynamic regions.
These transition times are the arrival time of a fluid front, displacing the low-attenuating oil phase by the high-attenuating brine phase. \\

\begin{figure}[!ht]
    \centering
    \includegraphics[width = \linewidth]{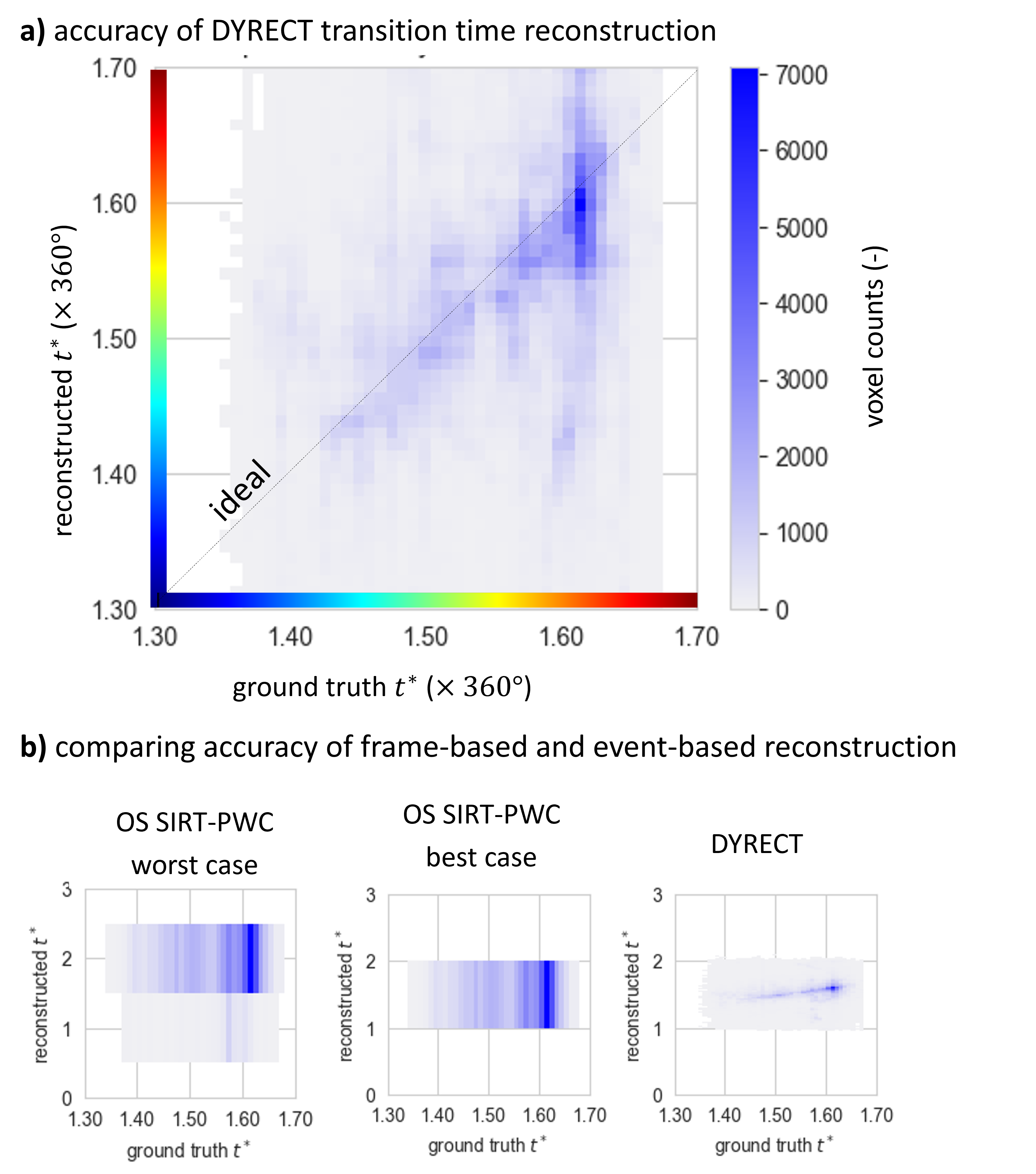}
    \caption{
    The 2D histogram (\textbf{a}) shows the voxel counts in the dynamic regions for each combination of ground truth transition time and the corresponding reconstructed value. The colour maps are equal to those used in figure~\ref{fig:rec_simulation_overlay}. Since most values are on the diagonal, this indicates that the DYRECT reconstruction of the 4D CT simulated dataset is able to retrieve the ground truth values with temporal deviations below a single rotation.
    The mean average error is 0.088\,CT rotations.
    Compared to a frame-based reconstruction in \textbf{b}, the mean absolute error is 0.076\,CT rotations in the best case and 0.460\,CT rotations in the worst case.
    }
    \label{fig:rec_simulation_histograms}
\end{figure}

The temporal resolution of the DYRECT reconstruction method is indicated by the histogram of co-occurrence between the transition times of the phantom and reconstruction (figure~\ref{fig:rec_simulation_histograms}a).
If the direct method succeeds in achieving a projection-level time resolution, transition times in a software phantom and those in the final reconstruction need to be a near-perfect mapping. This should be evident from their values following a diagonal in a co-occurrence histogram.
This deviation from the diagonal was quantified by the mean absolute error between the ground truth and the reconstruction (0.088\,CT rotations).
The time associated with this deviation corresponds to less than a tenth of the time required to record the projections for CT frames in a circular cone beam scan.
As a reference for this simulation study, we compared to a a frame-based approach here named OS SIRT-PWC (static reconstructions followed by a piecewise constant reduction of the temporal domain, reminiscent of~\cite{van2015iterative}), the accuracy depends on the coincidence of the time-frames with the ground truth events. That is because each frame corresponds to an interval in time needed for a full 360\textdegree\,rotation. The coarse temporal resolution of SIRT-PWC, as a frame-based technique, is illustrated through the vertical spread of the estimated event time (in blue) that is indicative of low precision.
For this method, the mean absolute errors in the best and worst case were 0.076 and 0.460\,CT rotations.
Since the main trend line was well matched in DYRECT (close to the ideal line in figure~\ref{fig:rec_simulation_histograms}a), this comparison with the frame-based method shows that future work exists in mitigating the outliers, for example by spatial regularization of the transition time during the reconstruction.
\\

\subsection{Angular dependency of temporal accuracy on direction of flow field}
\label{ssec:res:angular_flow}

\begin{figure}[!ht]
    \centering
    \includegraphics[width = \linewidth]{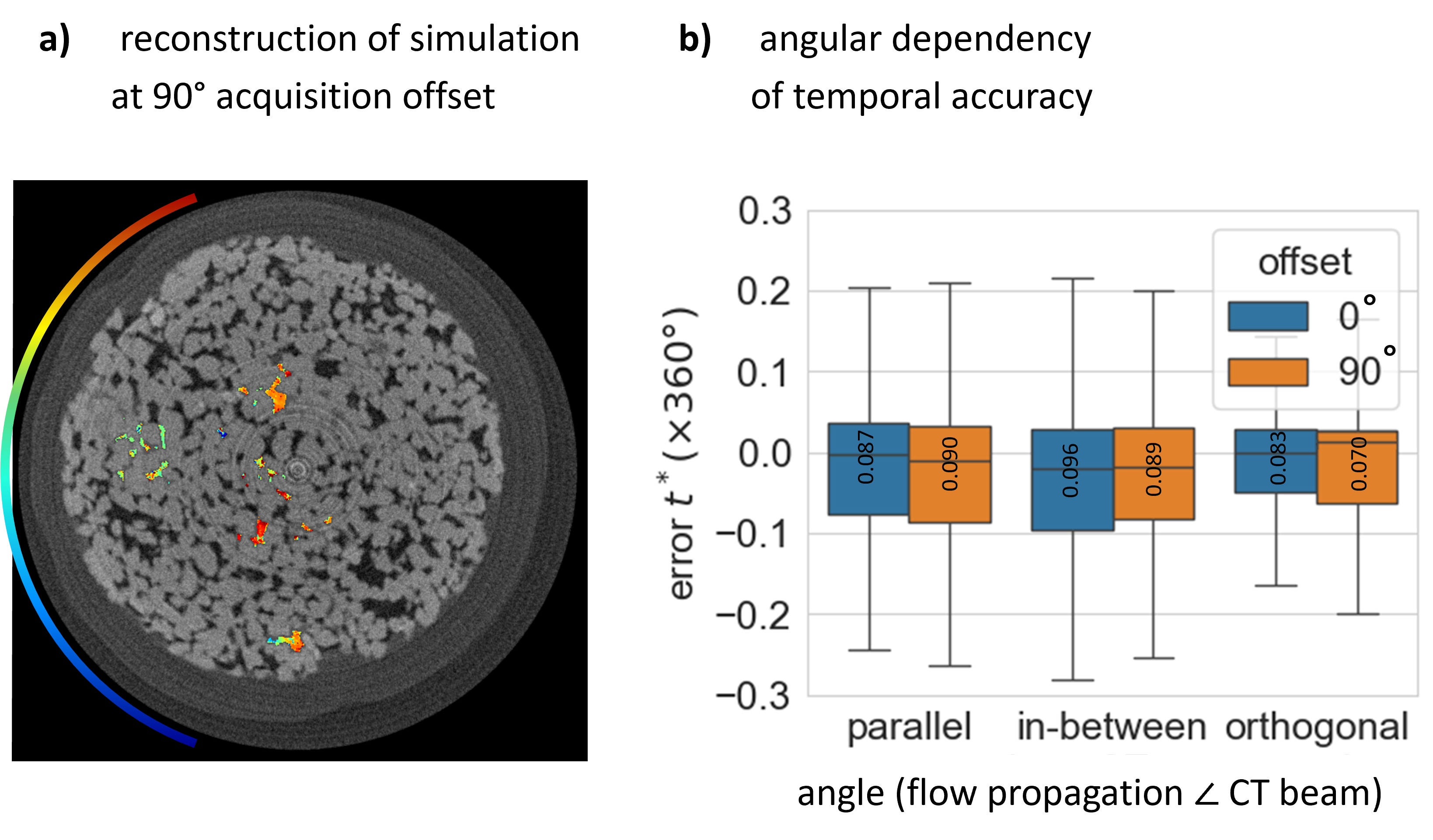}
    \caption{The instantaneous angle between the X-ray beam and the internal propagation direction may affect how accurately the transition time $t^*$ can be reconstructed. A second simulation (\textbf{a}), with equal dynamic conditions but offset by a 90\textdegree\,acquisition angle, yielded similar reconstruction quality (\textbf{a}) as the result of figure~\ref{fig:rec_simulation_overlay} with 0\textdegree\,offset. The mean absolute error on $t^*$ (\textbf{b}) overlaid on the boxplots was computed per instantaneous relative angle between the propagation direction and the optical axis vectors, categorised as parallel, orthogonal or in-between.
    }
    \label{fig:accuracy_angle}
\end{figure}

The uncertainty resulting from the complex interplay between acquisition direction and flow direction $\nabla_\mathbf{x} t^*(\mathbf{x})$ was further investigated in a second simulation study. 
Figure~\ref{fig:accuracy_angle} shows the reconstruction result for the simulation performed in figure~\ref{fig:rec_simulation_overlay}, but with a 90\textdegree\,offset in the acquisition angle (indicated by the coloured arc in sub-figure \textbf{a} that indicates the time-dependent viewing direction of the circular cone-beam CT system through the sample).
Apart from this, the dynamic ground truth sample remained exactly the same.
This examines whether the achieved temporal accuracy, measured in the simulation study, resulted from coincidental (mis-)alignment with the dynamic regions. 
Sub-figure~\ref{fig:accuracy_angle}\textbf{a} shows that the temporal accuracy was similar to the original outcome. This indicates that, for this simulated sample, there was no clear influence of the structural orientation. 
Similarly, the dependency on the angle between flow propagation and the instantaneous beam direction was studied in sub-figure~\ref{fig:accuracy_angle}b for angular categories parallel (0\textdegree\,to 20\textdegree\, \& 160\textdegree\,to 180\textdegree), near-orthogonal vectors (80\textdegree\,to 100\textdegree) and angles in-between.
The mean absolute errors and the boxplots show that movement orthogonal to the X-ray beam was reconstructed with a temporal accuracy ($0.070 - 0.083$ rotation periods or $25\degree - 29\degree$) that was marginally better than the accuracy for movement parallel to the X-ray beam ($0.087 - 0.090$ rotation periods or $31\degree-32\degree$). 
This insignificant difference could indicate that the sample was rotated fast enough so the simulated dynamics in the sandstone sample did not have a substantial impact. This leads to the recommendation that, when there is no clear structural anisotropy, the sample should be rotated fast and continuously at a constant rate to distinguish structures that possibly propagate in the direction of the optical axis. 
In both cases, the achieved temporal accuracy is better than conventionally achieved with one time frame per rotation, as indicated above.
\\

\section{Discussion}
\label{sec:discussion}

\subsection{Presence of acquisition noise and reconstruction artefacts}
In the reconstructions, the transition time contained non-physical noise. 
For instance, the fluid flow dataset was simulated with little intra-pore variations, and was still reconstructed with high variations within individual pores. In regions with a low attenuation difference between the initial and final state, this has little impact on the data fit with the projections.
Therefore, a low contrast to noise ratio leads to a decrease in certainty, which should be considered when the reconstruction is analysed for local displacements.
For instance, to reduce the impact of noise in the reconstruction, physical constraints could be incorporated in the form of spatial regularisation of the transition time volume.
This has not been done in the presented proof-of-concept reconstructions. 
The stability of the outcome is expected to be better when spatial correlations are incorporated into the physical model, e.g. stating that all voxels of a pore should change phase simultaneously or prior knowledge that there are distinct grey levels for static and dynamic regions~\cite{myers2011dynamic,batenburg2011dart}.
This comes at the cost of a bias towards more simplified dynamics, potentially missing finer details.
The analysis of the experimental dataset and the simulation showed that the proposed method already produces accurate results even without any spatial motion model, but the concept of the transition time volume is not restricted to this proposed optimisation strategy.
\\



\subsection{Implications of the single-step model}
The single-step model implies that there is temporal sparsity in the changes that occur within the sample.
Compared to frame-based methods that regularise by e.g. penalising temporal total variation~\cite{chen2009temporal,goethals2022dynamic}, this is done explicitly in the model.
This sparsity can be observed from an analysis point of view, when the experiment targets specific types of dynamics that happen under controlled conditions. 
However, even in those conditions, the model can fail to capture the changes in the sample. Consider for example a heterogeneous feature moving past a sampled position within the object. The evolution of the attenuation then corresponds to the streakline profile through the feature that moved through the object. 
This indicates that there is a relationship between the spatial texture of the object and the type of evolution that should be modelled within the sample.
In case the sample goes through multiple global stages of dynamics, the model in section~\ref{ssec:meth:rec} could be extended to support multiple steps by defining more global \textit{key frames} $\bm{\mu}_\tau$ (like $\bm{\mu}_A$ and $\bm{\mu}_B$) and transition time volumes $\mathbf{t}_{\tau}^*$, where the representative times of the key frames are chosen in advance.
To describe deformations of heterogeneous textures, projection-based displacement estimation methods can be used~\cite{jailin2018dynamic} instead of this fixed-coordinate representation.
However, explicit motion descriptions for dynamic CT reconstruction are generally made for samples with smooth deformation fields that can be decomposed into temporal and spatial dependencies.
\\

Defining the transition times on a continuous time scale theoretically allows much greater temporal resolution than post-reconstruction frame-based techniques.
To optimally resolve dynamics in the horizontal plane, the ideal acquisition still has a fast rotation during the acquisition of the subsequent projections. 
The issue of motion along the optical axis might be overcome better by multi-source systems that acquire projections simultaneously from at least two angles to leverage both the depth perception and the temporal resolution. 
Such examples are dual-source setups~\cite{maire2024dual}, a plenoptic imaging setup~\cite{longo2022flexible} or a swinging multi-source system~\cite{wu2017swinging}.
In turn, these different CT systems would require adaptations to the update strategy of the transition times like instantaneous triangulation.
That adaptation is needed since the symmetry of the subsequent 360\textdegree\,rotations was exploited strongly in the presented reconstruction technique.
Considering these future extensions, the main development that generally enabled a higher temporal resolution at a feasible data rate is the local event-based definition of the transition times. \\

Finally, since the simulation used the same parametrization of the dynamic volume as the forward projection operator of the proposed reconstruction technique, it was important to consider the possibility of an inverse crime~\cite{kaipio2007statistical}.
Avoiding this bias means that the ground truth and reconstruction should not use the same coordinate grid to represent the data, to avoid coincidentally good results that do not represent the typically expected results.
For instance, this could be done by simulating the ground truth on a rotated grid.
While this notion is relevant in validating spatial effects of reconstruction methods at high resolution, we are not yet at the single projection level in terms of temporal accuracy.
Therefore, the influence of discretization errors was expected to be minimal. \\


\section{Conclusion}
\label{sec:conclusion}

To overcome the issues related to dynamic CT imaging, namely the limited temporal resolution and the large data loads, a novel event-based reconstruction method was developed.
This study opens pathways to a new branch of dynamic CT reconstruction techniques that are both performant and data-efficient - without producing a global time-lapse of 3D frames. \\

The simulation study in section~\ref{ssec:res:simulated} indicated a high temporal accuracy, with a mean absolute error of 0.088\,CT rotations. 
The associated acquisition time corresponds to less than a tenth of the time required to record the 360\textdegree\,of projections for $\mu$CT frames in a circular cone beam scan with continuous acquisition.
Separating between movement in directions orthogonal and parallel with respect to the X-ray beam at the moment of transition, there was no sign that the studied scan configuration already suffered from inaccuracies for material propagation parallel to the X-ray beam.
Therefore, based on the current distinction between radioscopy and tomoscopy, this first implementation of the event-based reconstruction provides a promising outlook to achieve even higher gains in temporal resolution. 
Regarding reconstruction noise, better stability in the volume of transition times is expected from incorporating spatial prior knowledge in the iterative reconstruction scheme. \\

The work here puts a new perspective on the limits on temporal resolution that is practically achievable in dynamic CT imaging.
Usually, the gain in temporal resolution is quantified by the reduction in projections that are required per time frame to still reconstruct an accurate 3D sample representation. This relates to the spatial complexity of the structure, ignoring the temporal redundancies in longer dynamic CT acquisitions. 
This work has demonstrated that the temporal resolution of the reconstruction can be improved, depending on the complexity of the dynamics, and is to a small extent affected by the interplay between the CT acquisition angle and the propagation direction of the dynamic structures that undergo local changes in the attenuation coefficient. 
Whereas, previously, the rotation speed of the experiment needed to approach the temporal resolution of single events in frame-based reconstruction, it now becomes related to the time between subsequent events \textit{at the same location}. This relaxes the need for fast rotations to a degree of global object states instead of each individual local event. \\


The results on the experimental and synthetic dataset have shown that this reconstruction technique can accurately describe a variety of dynamic CT cases. 
This is required to experimentally quantify the phenomena that can make more efficient and durable processes and materials in industry and academia.
This reconstruction technique will enable researchers in many different research domains to delve deeper into their acquired $\mu$CT scans, resolving dynamics that previously remained unobserved due to the lack of temporal resolution and the time required to process the large 4D datasets.


\appendices

\section{Mitigating field of view effects and global instabilities on the experimental dataset}
In realistic experiments, additional factors, such as field of view and sample instability, may affect the accuracy of the reconstruction.
A first obstruction for the iterative reconstruction technique was that the field of view was too narrow to fully cover the surrounding crucible. 
Region of interest scanning poses a challenge to iterative reconstruction techniques, since the forward projection needs to simulate projections based on information that is not entirely covered in the field of view.
If this is not compensated for, the external mass is carried to the reconstruction space with an additional temporal dependency. 
Therefore, an initial reconstruction was made to estimate the density and position of the surrounding crucible. The resolved part of this cylinder was extrapolated homogeneously over all azimuthal angles.
The empty cylinder was projected in the scanned acquisition geometry to normalise all projections for the unresolved crucible.
\\

A second challenging factor was that, at higher dynamic CT acquisition rates, the experimental sample stability became less precise than typically encountered in static CT imaging.
An initial sliding window SART reconstruction showed that the sample was out of centre and oscillated over the x axis (which corresponds to the horizontal transverse direction in the sample with respect to the optical axis at 0\textdegree) with a period equal to the duration of 1 rotation.
This was compensated for by using the global affine motion compensation, summarised in table~\ref{tab:motion_models} in appendix~\ref{sec:app_motion}. 
\\

\section{Motion models}
\label{sec:app_motion}

With the addition of the event-based reconstruction, there are three major models to describe fast motion of the sample during the acquisition. 
The reconstruction of the experimental dataset captures a combination of global sample instabilities and local events.
These are listed in table~\ref{tab:motion_models}. 
A comparison between the three techniques was outside the scope of this study, the aim of this brief appendix is to guide the choice for (combinations of) motion models.
Since these models are capable of describing the same physical motion in a redundant way, it is important to maintain a clear hierarchy in updating the parameters of each model. 
\begin{enumerate}
    \item Affine coordinate transformations express global instabilities of the sample. This is a lightweight representation that remains constant in parameter count with improved spatial resolution. Accuracy can be improved by increasing the number of temporal control poses, and by using a physical model to interpolate the motion, e.g. a steady affine motion (SAM) model~\cite{rossignac2011steady}. 
    \item Digital Volume Correlation (DVC) is often used to estimate local sample deformations between two 3D frames at times $t_0$ and $t_1$ and to express these as a displacement vector field. The evolution over the projection acquisition time $t_p$ is usually expressed by a linear magnitude change from the reference time $t_r$~\cite{de2018motion}. The deformation field is not necessarily invertible~\cite{chen2008simple,van2017movit}, which could lead to issues in samples that display rapid accelerations, direction changes and spatially discontinuous motion fields as observed in fluid flow processes~\cite{bultreys2016fast}. 
    \item The spatial map of transition event times, proposed in this manuscript, captures irregular motion of homogeneous components. Additionally, dissolution and precipitation at the boundaries of structures with heterogeneous texture can be represented. 
    That is why the surface motion in a two-phase fluid flow experiment can be interpreted in relationship to $\nabla_\mathbf{x} t^*(\mathbf{x})$.
    If there is bulk material moving inside the sample, the texture contrast of that bulk should be negligible to the structural contrast between the materials. If this is not the case, motion artefacts will invalidate the temporal model of the step function in each individual voxel. 
\end{enumerate}

\begin{table}[t]
\begin{center}
    \caption{Three methods to address different aspects of sample motion in CT reconstruction. Motion can be modelled explicitly by an affine model or a displacement vector field, or implicitly by the event-based representation.}
    \resizebox{\linewidth}{!}{%
    \begin{tabular*}{1.3\linewidth}{cccc} 
    \hline
    \hline
    Dynamics & Global instabilities & Local displacement & Local events\\
    \arrayrulecolor{lightgray}\hline
    Motion model & SAM & Displacement vector field & Local phase change \\
    \hline
    Expression & $M_{0,1}\left(t_p-t_r\right)(\mathbf{x})$ & $M_{0,1}\left(t_p - t_r\right)(\mathbf{x})$ & $\mu(\mathbf{x},t_p)$\\
    & $ = A^{\dfrac{t_p-t_r}{t_1-t_0}} \cdot \mathbf{x}$ & $= \mathbf{x} + \mathrm{DVF}(\mathbf{x}) \dfrac{t_p - t_r}{t_1 - t_0}$ & $= \mu_0(\mathbf{x}) s(t^*(\mathbf{x}) - t_p)$\\
    & & & $+ \mu_1(\mathbf{x}) s(t_p - t^*(\mathbf{x}))$\\
    \hline
    Motion evolution & Steady affine & Linear local direction & General \\\hline
    Moving structures & General texture & General texture & Homogeneous texture\\
    \hline
    Fixed structures & Static & Static & \begin{tabular}{c}
                                          Dissolution \\
                                          Precipitation
                                          \end{tabular}\\
    \hline
    Parameter count & $O(N_t)$ & $O(N_\mathbf{x} \times N_t)$ & $O(N_\mathbf{x})$ \\
    \arrayrulecolor{black}
    \hline
    \hline
    \end{tabular*}
    }
    \label{tab:motion_models}  
\end{center}
\end{table}

\section*{Acknowledgements}
Dr. Paul Hans Kamm is gratefully acknowledged for providing the CT data and additional information of the bubble coalescence experiment.
Ghent University Special Research Fund (BOF-UGent) is acknowledged for the support to the UGCT Core Facility (BOF.COR.2022.0009).
This work is supported by ERC grant 101116228 / Flowscopy. 
Funded by the European Union. Views and opinions expressed are however those of the author(s) only and do not necessarily reflect those of the European Union or the European Research Council Executive Agency. Neither the European Union nor the granting authority can be held responsible for them.
Shell Global Solutions International B.V. is acknowledged for financial and technical support of this work, and providing the sandstone material and dynamic $\mu$CT scan that were used as a basis for the simulation study.
The authors have no competing interests to declare that are relevant to the content of this article. \\

\section*{Data availability}
The full experimental dataset of bubble coalescence is available on \url{https://doi.org/10.16907/ d7582cb6-7850-42bc-ad76-e845b998e9ca} in dataset in Dataset 20.500.11935/6285a066-8d02-4b7e-94f6-4f2efcdcb8af.
The reconstructions of the simulated and experimental dataset can be accessed on
\url{https://osf.io/64erh/?view_only=a5881a56118b4fb783c29de05365fb4b}
The core GPU code and metadata management for DYRECT are available in supplementary files. Supporting code of CTrex is available with the author upon request.



\bibliographystyle{IEEEtran}
\bibliography{citations}

 



\vfill

\end{document}